\documentclass{aa}
\usepackage{graphicx}
\begin{document}

\title{Ionization structure in the winds of B[e] supergiants}
\subtitle{I. Ionization equilibrium calculations in a H plus He wind}

\author{M. Kraus
  \and
   H.J.G.L.M. Lamers 
  }

\offprints{M. Kraus}

\institute{Astronomical Institute, University of Utrecht, Princetonplein 5, NL-3584 CC Utrecht\\
           \email{M.Kraus@phys.uu.nl; lamers@astro.uu.nl} 
      }

\date{Received; accepted }

\abstract{

The non-spherically symmetric winds of B[e] supergiants are investigated. An empirical 
density distribution is chosen that accounts for the density concentrations and ratios
derived from observations, and our model winds are assumed to contain only hydrogen and 
helium.
We first calculate the approximate ionization radii for H and He and compare the
results with the ionization fractions calculated from the more accurate ionization balance 
equations. We find that winds with a $r^{-2}$ density distribution turn out to reach a 
constant ionization fraction as long as the wind density is low, i.e. in polar direction. For 
the high density equatorial regions, however, we find that the winds become neutral just 
above the stellar surface of the hot and massive B[e] supergiants forming a disk-like neutral 
region. In such a disk molecules and dust can form even very near the hot central star.


}

\maketitle

\section{Introduction}

B[e] supergiants are massive and luminous post-main sequence stars with strong non-spherically
symmetric winds induced probably due to the rapid rotation of the stars. Observations of dust 
and molecules indicate that there must exist some neutral regions near these stars that are 
shielded from the ionizing and dissociating stellar radiation field. The goal of this paper 
is to show that in the equatorial regions where the slow, high density wind forms a disk-like
structure, hydrogen becomes neutral just above the stellar surface and can therefore 
perfectly shield the disk material to allow molecules and dust to be formed near the
hot central star.

Photoionization of gas around hot stars is since long a well known process leading to
the beautiful observable emission objects like \ion{H}{ii} regions and planetary 
nebulae. The central stars thereby ionize the material that has been blown away by the
star itself in a more or less dense wind. The sizes of such ionization regions, also called
the Str\"{o}mgren spheres, can be calculated by balancing the number of ionizations by the
number of recombinations and usually a very sharp outer edge of the ionized region is found.
These calculations are normally performed for nebulae with a constant electron temperature
and density (for an overview see Osterbrock (\cite{osterbrock}) and references therein).

O and B stars have a stellar wind which for a spherically symmetric and steady 
radial outflow leads to a $v(r)^{-1}r^{-2}$ wind density distribution. So the calculation of 
the ionization structure in these winds is more complicated than for a
homogeneous cloud. In an early work by Drew 
(\cite{drew85}) a radiative ionization and thermal equilibrium model for the wind from 
P\,Cygni was calculated, and later on e.g. Pauldrach (\cite{pauldrach87}),
Drew (\cite{drew89}), MacFarlane et al. (\cite{macfarlane}), and Pauldrach et al. 
(\cite{pauldrach}) calculated in
great detail the ionization structure in the {\it spherically} symmetric winds of hot luminous
stars using the ionization balance equations.  In addition, they solved  the thermal equilibrium
equations to calculate the electron temperature in the wind. For this, line cooling by 
collisionaly excited heavy elements is the most important cooling mechanism and the authors
took a huge amount of different lines into account which results in an enormous set of 
equations that has to be solved.

In this paper we calculate the ionization structure in the non-spherically symmetric winds
of B[e] supergiants. For simplicity, these winds are supposed to contain only hydrogen and 
helium. We especially focus our investigations on the equatorial regions where
we expect to find neutral material near the central star. In our calculations we make use
of several approximations which result in an {\it upper limit} for the distance of the
equatorial neutral region, because we want to show that efficient shielding of
dust at larger distances can occur.

\section{The winds of B[e] supergiants}\label{winds}

B[e] supergiants are thought to be post main-sequence stars but their real location 
within a clear evolutionary sequence is still under debate (see e.g. Schulte-Ladbeck 
\cite{schulte-ladbeck}; Lamers et al. \cite{lamers03}). 
Up to now about 15 B[e] supergiants are known in the Magellanic Clouds, and a few candidates
are known in the Milky Way although the classification as a B[e] supergiant in the Milky Way 
is very difficult due to the lack of good distance and therefore luminosity determinations
(Lamers et al. \cite{lamers98}). The situation is more favorable for B[e] supergiants in the 
Magellanic Clouds for which the distances are known.

The B[e] supergiants in the Magellanic Clouds have been observed in great detail and many
spectra in different wavelength regions are nowadays available. Most interesting, these 
stars were found to have a so-called hybrid spectrum (Zickgraf et al. \cite{zickgraf85}) 
which means that on the one hand the UV resonance lines of highly ionized atoms show broad 
P\,Cygni profiles indicating a wind of low density and high velocity, while, on the other 
hand, forbidden and permitted emission lines of low ionized ions and of neutral atoms arising 
in the optical spectrum indicate a wind of high density and low velocity. The winds of B[e] 
supergiants must therefore consist of two different components and Zickgraf et al. 
(\cite{zickgraf85}) proposed a picture of these winds containing a hot and fast 
radiation-driven CAK-type wind (Castor et al. \cite{castor}) in polar direction and a much 
cooler, slow and dense wind in equatorial direction which might form a disk-like structure.
The density contrast between equator and pole is of order $10^{2} - 10^{3}$, and from
the statistical viewpoint a mean opening angle of $20\degr - 40\degr$ of the disk has 
been derived (Zickgraf \cite{zickgraf90}, \cite{zickgraf92}). The reason for the 
non-spherically symmetric winds is probably the rapid rotation of these stars. Rotational
velocities in the equatorial region of the star of about $40 - 70\,\%$ of the break-up 
velocity have been derived from their spectra (Zickgraf et al. \cite{zickgraf96}).
Pelupessy et al. (\cite{pelupessy}) and Bjorkman (\cite{bjorkman98}) have argued
that the disk is probably due to a combination of rotation induced bi-stability
and rotation induced wind compression (see Lamers \& Casinelli \cite{lamers-casinelli}, Chapter 11). 

Other observations confirm the existence of a neutral disk. These are emission from
molecules like CO (McGregor et al. \cite{mcgrega}, \cite{mcgregb}) and TiO (Zickgraf et al.
\cite{zickgraf89}) and emission from cool and warm dust with strong contributions to the 
spectral energy distribution (Zickgraf et al. \cite{zickgraf85}, \cite{zickgraf86}, 
\cite{zickgraf89}, \cite{zicketal92}; Gummersbach et al. \cite{gummers}). Also
some typical dust emission features have been observed (e.g. Waters et al. \cite{waters}; Allen \cite{allen}; Zickgraf
et al. \cite{zickgraf86}). The dust component in these disks is also confirmed via 
polarization measurements (Magalh$\tilde{\rm a}$es \cite{magalhaes}). Detailed observations 
of the B[e] supergiants with different inclination angles show that the polarization is
strongest for the edge-on systems. In addition, there is a correlation between the 
polarization and the [$H - L$] color of the stars, indicating that the observed infrared
excess emission is due to circumstellar dust concentrated in the equatorial region. This 
dust disk is the origin of the major part of the observed intrinsic polarization.

\section{The wind model}

\begin{figure}
\resizebox{\hsize}{!}{\includegraphics{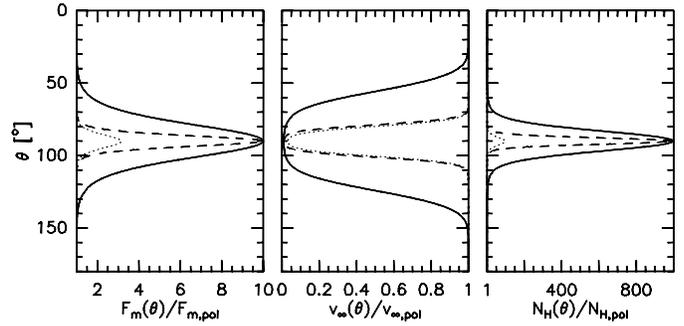}}
\caption{Angular distribution of the mass flux,
velocity and hydrogen density on the stellar surface. The parameters are
$b_{1} = 1; b_{2} = -2; s = 100$ (dashed lines);
$b_{1} = 0.5; b_{2} = -1.5; s = 100$ (dotted lines); and
$b_{1} = 1; b_{2} = -2; s = 10$ (solid lines).}
\label{ang}
\end{figure}          

For a star with a stationary spherically symmetric wind, the mass loss rate is related to the density and the velocity at any point in the wind via the equation of mass continuity
\begin{equation}\label{mdot_sphere}
\dot{M} = 4\pi r^{2} \rho(r)\,v(r)
\end{equation}
The mass flux per unit surface area from a spherical star is
\begin{equation}
F_{\rm m} = \frac{\dot{M}}{4\pi R_{*}^{2}} = \rho(R_{*})\,v(R_{*})
\end{equation}
For a non-spherically symmetric wind the mass flux is a function of latitude, $F_{\rm m} = 
F_{\rm m}(\theta)$,  
%
%
and the mass loss rate is
\begin{equation}\label{mdot_sector}
\dot{M} = 4\pi R_{*}^{2}\int\limits_{0}^{\pi/2}F_{\rm m}(\theta)\,\sin(\theta)\,d\theta
\end{equation}
where we assumed that the star is a sphere.
The hydrogen density distribution in such a wind is
\begin{equation}\label{dens} 
N_{\rm H}(\theta, r) = \frac{1}{\mu m_{\rm H}}\left(\frac{R_{*}}{r}\right)^{2} \frac{F_{\rm m}(\theta)}
{v(\theta,r)}
\end{equation}
which we parametrize in the following form
\begin{equation}\label{dens_param}
N_{\rm H}(\theta, r) = \frac{1}{\mu m_{\rm H}}\left(\frac{R_{*}}{r}\right)^{2}\frac{F_{\rm m, pol}}
{v_{\rm pol}(r)}\cdot 10^{b\cdot (\sin\theta)^{s}}
\end{equation}  
\begin{table*}
\caption{Model wind parameters. Listed are the electron temperature, $T_{\rm e}$,  
mass fluxes in polar and equatorial direction, $F_{\rm m, pol}$ and $F_{\rm m, eq}$,
mass loss rate of the star, $\dot{M}$, terminal velocity in equatorial direction, $v_{\infty, 
eq}$, and the resulting equatorial hydrogen density on the stellar surface, $N_{\rm H, eq}(R_{*})$.
The velocity in polar direction is assumed to be in all cases 2000\,km/s which is a typical 
value in a CAK-type wind.}
\begin{tabular}{lccccccrc}
\hline
\hline
Model & $T_{\rm e}$ & $F_{\rm m, pol}$ & $F_{\rm m, eq}$ & $\dot{M}$ & $v_{\infty, eq}$ & 
 $b$ & $s$ & $N_{\rm H, eq}(R_{*})$\\
 & [K] & [g s$^{-1}$ cm$^{-2}$] & [g s$^{-1}$ cm$^{-2}$] & [M$_{\sun}$/yr] & [km/s] & & & [cm$^{-3}$] \\
\hline
A & $1.0\cdot 10^{4}$ & $1.5\cdot 10^{-7}$ & $1.5\cdot 10^{-6}$ & $2.4\cdot 10^{-6}$ & 20 & 3.0 & 100 & $3.3\cdot 10^{11}$\\
B & $1.5\cdot 10^{4}$ & $1.5\cdot 10^{-7}$ & $1.5\cdot 10^{-6}$ & $2.4\cdot 10^{-6}$ & 20 & 3.0 & 100 & $3.3\cdot 10^{11}$\\
C & $1.0\cdot 10^{4}$ & $1.5\cdot 10^{-6}$ & $1.5\cdot 10^{-5}$ & $2.4\cdot 10^{-5}$ & 20 & 3.0 & 100 & $3.3\cdot 10^{12}$\\
D & $1.0\cdot 10^{4}$ & $1.5\cdot 10^{-8}$ & $1.5\cdot 10^{-7}$ & $2.4\cdot 10^{-7}$ & 20 & 3.0 & 100 & $3.3\cdot 10^{10}$\\
E & $1.0\cdot 10^{4}$ & $1.5\cdot 10^{-7}$ & $4.9\cdot 10^{-7}$ & $1.8\cdot 10^{-6}$ & 63 & 2.0 & 100 & $3.3\cdot 10^{10}$\\
F & $1.0\cdot 10^{4}$ & $1.5\cdot 10^{-7}$ & $1.5\cdot 10^{-6}$ & $4.0\cdot 10^{-6}$ & 20 & 3.0 &  10 & $3.3\cdot 10^{11}$\\
\hline
\end{tabular}
\label{windtab}
\end{table*} 
The term $10^{b}$ with $b = 2 - 3$ reflects the density contrast between the equator
and the pole, and the term $(\sin\theta)^{s}$ describes the stronger density concentration
with increasing $s$ towards the equator. The parameters $F_{\rm m, pol}$ and $v_{\rm pol}(r)$
are the mass flux and wind velocity in polar direction which can be derived from the
observations. Their latitude dependence is taken to
\begin{equation}
F_{\rm m}(\theta) = F_{\rm m, pol}\cdot 10^{b_{1}(\sin\theta)^{s}}
\end{equation}
and
\begin{equation}
v(\theta,r) = v_{\rm pol}(r)\cdot 10^{b_{2}(\sin\theta)^{s}}
\end{equation}
with $b = b_{1} - b_{2}$. This implies that we have adopted the same shape of
the velocity law, $v(r)/v_{\infty}$, for all latitudes.
In Fig.\,\ref{ang} we show the variation of the mass flux, velocity and hydrogen density
along the stellar surface for different values of $b_{1}, b_{2}$, and $s$.       

The radial velocity dependence in a stellar wind is normally described by the $\beta$-law
\begin{equation}\label{beta}
v(r) \simeq v_{\infty}\left( 1-\frac{R_{*}}{r}\right)^{\beta}
\end{equation}
where $v_{\infty}$ is the terminal velocity.
For hot stars, $\beta$ is typically of order 0.8, and the wind reaches its terminal velocity
at a distance of a few stellar radii. 

For our calculations we make the following assumptions, whose validity and influence on the
results are discussed in Sect.\,\ref{discuss}:\\
(i) we set the electron temperature in the wind constant,\\
(ii) we take the velocity in the wind to be independent of distance, i.e. $v(\theta,r) =
v_{\infty}(\theta)$.

For this paper we choose a star with a mass of approximately $M_{*} =
30$\,$M_{\sun}$, a radius of $R_{*} = 82$\,$R_{\sun}$, and a radiation temperature of
the photosphere in the Lyman continuum of $T_{\rm rad} = 1.7\cdot 10^{4}$\,K which 
coincides with a Kurucz model atmosphere (Kurucz \cite{kurucz}) for a star with an 
effective temperature $T_{\rm eff} \simeq 22\,500$\,K and $\log g = 3$. The
luminosity of this star is $L_{*} = 1.5\cdot 10^{6}$\,L$_{\odot}$ which
corresponds to the most luminous B[e] supergiants observed.
 
For the wind we calculate different models whose parameters are summarized in
Table\,\ref{windtab}.

\section{Ionization radii in a wind with non-spherical density distribution}

In this section we discuss the calculation of ionization radii. We adopt a similar
procedure as for the classical Str\"{o}mgren sphere calculation but use the non-spherical
density distribution given by Eq.\,(\ref{dens_param}). We first derive the general formulae. 
In Sect.\,\ref{thin} we discuss the ionization radii in the optically thin case while in 
Sect.\,\ref{thick} the on-the-spot approximation is used. Both models are with the assumption 
that the ionization is complete within the ionization radii (as usually adopted). This 
simplification allows us to get a first impression of the overall shape of the ionized region 
around a B[e] supergiant (Sect.\,\ref{result}). The physically more relevant case of partial 
ionization is discussed in Sect.\,\ref{ionstruc}.

\subsection{General formulae}

We restrict the calculations to a wind containing hydrogen and helium.
Since the following equations have the same shape for all angles
$\theta$ we skip the angular dependence of the densities and
write down the equations only as functions of the radial distance.
In addition, we assume that the fraction of He$^{++}$ is negligible.

\begin{figure}
\resizebox{\hsize}{!}{\includegraphics{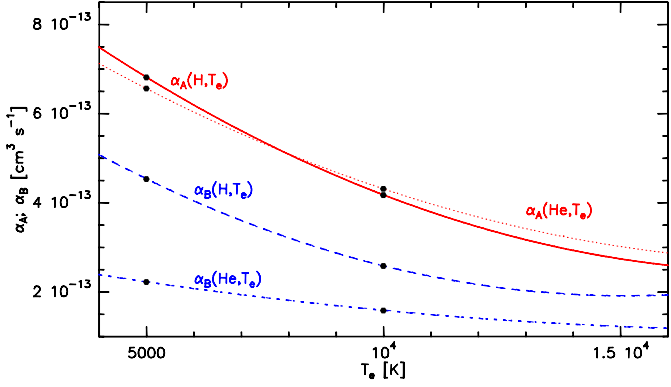}}
\caption{Hydrogen and helium recombination coefficients. The dots for H refer to values given
by Seaton (\cite{seaton}), the dots for He are taken from Osterbrock (\cite{osterbrock}), and
the lines give fits according to interpolation calculations.}
\label{rec}
\end{figure}

The ionization radius is calculated by balancing the number of ionizations which depends
an both, the stellar, $J_{\nu}^{\rm s}$, and the diffuse radiation field, $J_{\nu}^{\rm d}$, 
and the number of recombinations per cm$^{3}$ and per s. For H and He we can therefore write
\begin{eqnarray}\label{bal_H1}
N_{\rm H^{0}}(r) & \int\limits_{\nu_{0}}^{\infty} & a_{\nu}({\rm H^{0}})\left[
\frac{4\pi J_{\nu}^{\rm s}}{h\nu} + \frac{4\pi J_{\nu}^{\rm d}}{h\nu}\right]
 d\nu \nonumber \\
& = & N_{\rm H^{+}}(r)N_{\rm e}(r) \alpha_{\rm A}({\rm H^{0}},T)
\end{eqnarray}
\begin{eqnarray}\label{bal_He1}
N_{\rm He^{0}}(r) & \int\limits_{\nu_{2}}^{\infty} & a_{\nu}({\rm He^{0}})\left[
\frac{4\pi J_{\nu}^{\rm s}}{h\nu} + \frac{4\pi J_{\nu}^{\rm d}}{h\nu}\right]
 d\nu \nonumber \\
& = & N_{\rm He^{+}}(r)N_{\rm e}(r) \alpha_{\rm A}({\rm He^{0}},T)
\end{eqnarray}
with $N$ as the number density of the individual ions and neutral atoms, whereby $N_{\rm 
He} = y\cdot N_{\rm H}$ and $y$ is the He/H ratio by number which we set to 0.1. The terms
$\alpha_{\rm A}({\rm H^{0}},T)$ and $\alpha_{\rm A}({\rm He^{0}},T)$ are the radiative recombination
coefficients of hydrogen and helium for recombinations into all levels (see Fig.\,\ref{rec}), 
$a_{\nu}({\rm H^{0}})$ and $a_{\nu}({\rm He^{0}})$ are the ionization coefficients, $\nu_{0}$
and $\nu_{2}$ are the ionization frequencies of hydrogen and helium, and the electron 
density, $N_{\rm e}$, is given by $N_{\rm e}(r) = N_{\rm H^{+}}(r) + N_{\rm He^{+}}(r)$.

Taking into account the absorption of stellar continuum photons along the direction 
to the star, the mean intensity of the stellar radiation is  
\begin{equation}\label{stellar}
4\pi J_{\nu}^{\rm s} = 4\pi W(r) B_{\nu}(T_{\rm rad}) e^{-\tau_{\nu}}
\end{equation}
where $W(r) = \frac{1}{2}(1-\sqrt{1 - (R_{*}/r)^{2}})$ is the geometrical dilution 
factor, $T_{\rm rad}$ is the radiation temperature of the photosphere, and 
$\tau_{\nu}$ is the radial optical depth defined as
%
%
\begin{equation}\label{tau1}
\tau_{\nu}(r) = \int\limits_{R_{*}}^{r}a_{\nu}({\rm H^{0}})N_{\rm H^{0}}(r') dr'
\end{equation}
for $\nu_{0} \leq \nu < \nu_{2}$ and
\begin{equation}\label{tau2}
\tau_{\nu}(r) = \int\limits_{R_{*}}^{r}\left( a_{\nu}({\rm H^{0}})N_{\rm H^{0}}(r') +
a_{\nu}({\rm He^{0}})N_{\rm He^{0}}(r')\right) dr'
\end{equation}
for $\nu_{2} \leq \nu$. 
We know that Eq.\,(\ref{stellar}) is only correct as long as the star can be considered as
a point source. But for our purpose this approximation is also good enough for distances 
near the star (see Sect.\,\ref{discuss}).

The diffuse radiation field is much more complicated to calculate because at every point
in the wind one must integrate the diffuse radiation over the complete wind zone. 
To do so, one needs the information of the structure of the wind zone, i.e. information about
where in the non-spherically symmetric wind exist neutral regions, what is the 
absorptivity and emissivity of these neutral regions, etc.
Therefore, we restrict our investigations to two extremes: (a) the optically thin case,
where all ionizing radiation produced by recombinations leaves the wind zone without being
absorbed by a neutral atom, and (b) the optically thick case, better known as the on-the-spot
approximation, which states that every ionizing photon produced in the wind via recombination
is immediately absorbed by a neutral atom nearby, i.e. on-the spot. In reality the mean
intensity will be in between the values predicted for these two extremes.

\subsection{Non-spherical ionization radii in the optically thin approximation}\label{thin}

In this case, the diffuse radiation field is zero, and the ionization equilibrium equations
for H and He, Eqs.\,(\ref{bal_H1}) and (\ref{bal_He1}), reduce to
\begin{eqnarray}\label{rs_H_thin}
4\pi W(r) (1  -  q_{\rm H}(r))\int\limits_{\nu_{0}}^{\infty}\frac{B_{\nu}(T_{\rm 
rad})a_{\nu}({\rm H^{0}})}{h\nu}\,e^{-\tau_{\nu}(r)}\,d\nu \nonumber \\
  =  q_{\rm H}(r)\,N_{\rm H}(r)\,(q_{\rm H}(r) + yq_{\rm He}(r))\,\alpha_{\rm A}({\rm H^{0}},T) 
\end{eqnarray}
and
\begin{eqnarray}\label{rs_He_thin}
4\pi W(r) (1  -  q_{\rm He}(r))\int\limits_{\nu_{2}}^{\infty}\frac{B_{\nu}(T_{\rm
rad})a_{\nu}({\rm He^{0}})}{h\nu}\,e^{-\tau_{\nu}(r)}\,d\nu \nonumber \\
 = q_{\rm He}(r)\,N_{\rm H}(r)\,(q_{\rm H}(r) + yq_{\rm He}(r))\,\alpha_{\rm A}({\rm He^{0}},T)
\end{eqnarray}
where we replaced the densities by density fractions $N_{\rm H^{+}}(r) = q_{\rm H}(r)\cdot 
N_{\rm H}(r)$ and $N_{\rm H^{0}}(r) = (1 - q_{\rm H}(r))\cdot N_{\rm H}(r)$, and the same 
for He.

To calculate the distance from the star at which the gas turns from ionized into neutral, 
i.e. the ionization radius $R_{\rm s}(\theta)$, we assume (as usually adopted in classical 
Str\"{o}mgren sphere calculations) that all atoms are ionized whithin 
the ionization radius, and all atoms are neutral outside, which means that
\begin{equation}
q_{\rm H} =
\left\{\begin{array}{lcl}
1 & ; & r < R_{\rm s}^{\rm H} \\
0 & ; & r > R_{\rm s}^{\rm H}\end{array}\right. 
\quad {\rm and} \quad
q_{\rm He} =
\left\{\begin{array}{lcl}
1 & ; & r < R_{\rm s}^{\rm He}\\
0 & ; & r > R_{\rm s}^{\rm He}\end{array}\right.
\label{q_define} 
\end{equation}
Integrating now both sides of Eqs.\,(\ref{rs_H_thin}) and (\ref{rs_He_thin}) over 
the radius, and taking into account the different sizes of the ionized regions for H and He
with $R_{\rm s}^{\rm He} \leq R_{\rm s}^{\rm H}$,
the ionization balance equations become 
\begin{eqnarray}\label{hydro}
4\pi \int\limits_{R_{\rm s}^{\rm H}}^{\infty}\!\!W(r)\int\limits_{\nu_{0}}^{\infty}\frac{B_{\nu}
a_{\nu}({\rm H^{0}})}{h\nu}\,e^{-\tau_{\nu}(r)}\,d\nu\,dr \qquad\qquad\qquad\nonumber \\
= \alpha_{\rm A}({\rm H^{0}},T)\left[\int\limits_{R_{*}}^{R_{\rm s}^{\rm He}}\!\!N_{\rm H}(r)(1 + y)\,dr
+\!\!\int\limits_{R_{\rm s}^{\rm He}}^{R_{\rm s}^{\rm H}}\!\!N_{\rm H}(r)\,dr \right]
\end{eqnarray}
for hydrogen, and
\begin{eqnarray}\label{heli}
4\pi \int\limits_{R_{\rm s}^{\rm He}}^{\infty}\!\!W(r)\int\limits_{\nu_{2}}^{\infty}\frac{B_{\nu}
a_{\nu}({\rm He^{0}})}{h\nu}\,e^{-\tau_{\nu}(r)}\,d\nu\,dr \nonumber \\
= \alpha_{\rm A}({\rm He^{0}},T)\int\limits_{R_{*}}^{R_{\rm s}^{\rm He}}\!\!N_{\rm H}(r)(1 + y)\,dr
\end{eqnarray}
for helium. The right-hand side of Eq.\,(\ref{hydro}) splits into two integrals according to 
the two different regions that have to be considered: (i) the region between the stellar 
surface and the He ionization radius where both, H and He are singly ionized, and 
(ii) the region between the He and H ionization radii where only H is ionized. 
Obviously, the corresponding integration limits for the optical depth calculation must be 
adapted correctly. From Eqs.\,(\ref{hydro}) and (\ref{heli}) the ionization radii for H, 
$R_{\rm s}^{\rm H}$, and He, $R_{\rm s}^{\rm He}$, can be calculated for every angle $\theta$.

These equations look quite different from the classical Str\"{o}mgren sphere 
formula. But since they are more general, the classical Str\"{o}mgren sphere,
as a special solution, can be found easily as shown in Appendix\,A.

\subsection{Non-spherical ionization radii in the on-the-spot approximation}\label{thick}

The on-the-spot approximation for optically thick winds assumes that every ionizing photon 
generated by recombination will be absorbed very near its generation location. The diffuse 
field can be split into several parts and we consider diffuse photons generated by (a) 
recombining hydrogen atoms into the ground level, (b) recombining helium atoms into excited 
levels, and (c) recombining helium atoms into the ground level.  
In this paper we take into account neither ionization from excited levels nor collisional
ionization and recombination. Both effects will strongly influence the ionization balance but we
expect that they become really important only for stars with much higher radiation temperature in the Lyman continuum 
than the 17\,000\,K we are dealing with. A full solution of the level population equations 
together with the ionization balance equations, where these effects are included, is presently 
under investigation and will be published in a forthcoming paper.  

While all of the diffuse photons generated by the three mentioned processes are able to ionize 
hydrogen, only photons resulting from the third process can ionize helium. To complete the 
ionization equilibrium equations for H and He in the on-the-spot approximation, we have to 
add the term 
\begin{eqnarray}\label{H_ots}
N_{\rm H}(r)~(q_{\rm H}(r) + yq_{\rm He}(r))~[\,q_{\rm H}(r)\,\alpha_{1}({\rm H^{0}},T) 
   \nonumber \\
+ yq_{\rm He}(r)\,(\xi \,\alpha_{1}({\rm He^{0}},T) + p\,\alpha_{\rm B}({\rm He^{0}},T))]
\end{eqnarray}
that describes the additional ionizations of H due to the diffuse
radiation field, to the left-hand side of Eq.\,(\ref{rs_H_thin}), and the term
\begin{equation}\label{He_ots}
N_{\rm H}(r)\,(q_{\rm H}(r) + yq_{\rm He}(r))\,yq_{\rm He}(r)\,(1-\xi)\,\alpha_{1}({\rm 
He^{0}},T)
\end{equation}
for the additional ionization of He due to the diffuse radiation field, to the left-hand side of
Eq.\,(\ref{rs_He_thin}). With the electron density in terms of the $q$-values,
$N_{\rm e}(r) = N_{\rm H}(r)\,(q_{\rm H}(r) + yq_{\rm He}(r))$, the ionization balance 
equations that must now be solved can be written in the following form:
\begin{eqnarray}\label{H_ots2}
4\pi W(r)\,(1  -  q_{\rm H}(r))\int\limits_{\nu_{0}}^{\infty}\frac{B_{\nu}(T_{\rm
rad})\,a_{\nu}({\rm H^{0}})}{h\nu}~e^{-\tau_{\nu}(r)}\,d\nu~\nonumber \\
+~N_{\rm e}(r)\left[\,q_{\rm H}(r)\,\alpha_{1}({\rm H^{0}},T)+ y\,q_{\rm 
He}(r)\,(\xi\,\alpha_{1}({\rm He^{0}},T)\right. \nonumber \\
\left.+~p\,\alpha_{\rm B}({\rm He^{0}},T))\,\right]~\!=~ N_{\rm e}(r)~q_{\rm 
H}(r)\,\alpha_{\rm A}({\rm H^{0}},T)\qquad\, 
\end{eqnarray}
for hydrogen, and
\begin{eqnarray}\label{He_ots2}
4\pi W(r)\,(1 - q_{\rm He}(r))\!\!\int\limits_{\nu_{2}}^{\infty}\frac{B_{\nu}(T_{\rm
rad})a_{\nu}({\rm He^{0}})}{h\nu}\,e^{-\tau_{\nu}(r)}\,d\nu + N_{\rm e}(r) \nonumber \\ 
\cdot\,y\,q_{\rm He}(r)\,(1-\xi)\,\alpha_{1}({\rm He^{0}},T) = N_{\rm e}(r) q_{\rm 
He}(r) \alpha_{\rm A}({\rm He^{0}},T)~~~~~
\end{eqnarray}
for Helium. Here, $\alpha_{\rm B}$ is the recombination coefficient to all levels except of the 
ground level, i.e. $\alpha_{\rm B} = \alpha_{\rm A} - \alpha_{1}$, and $\xi$ is the fraction of 
recombining helium to the ground level which is absorbed by H atoms. The remaining fraction 
$1 - \xi$ is absorbed by He atoms. $p$ is the fraction of photons generated by recombination 
of He to all excited levels that can ionize H. This fraction strongly depends on the electron 
density, because, if the electron density, $N_{\rm e}(r)$, at the location $r$ is high enough,
then collisions of electrons with the excited He atoms will depopulate the levels and no
or only a few photons are produced. The critical density, $N_{\rm c}$, for collisional 
deexcitation is defined as the density at which collisional transitions are equally probable 
with radiative transitions. The values for the ratio $p$ are either $p \simeq 0.96$ if 
$N_{\rm e}(r) \ll N_{\rm c}(r)$ or $p\simeq 0.66$ if $N_{\rm e}(r)\gg N_{\rm c}(r)$ (see
Osterbrock \cite{osterbrock}).
   
Making again the assumptions that all atoms are ionized within and neutral outside the
ionization radius, ignoring the absorption by H in the He$^{+}$ zone by setting $\xi = 0$,
and integrating the complete ionization equilibrium equations over the radius yields again
$R_{\rm s}(\theta)$ for H and He.

\subsection{Results}\label{result}

\begin{figure}
\resizebox{\hsize}{!}{\includegraphics{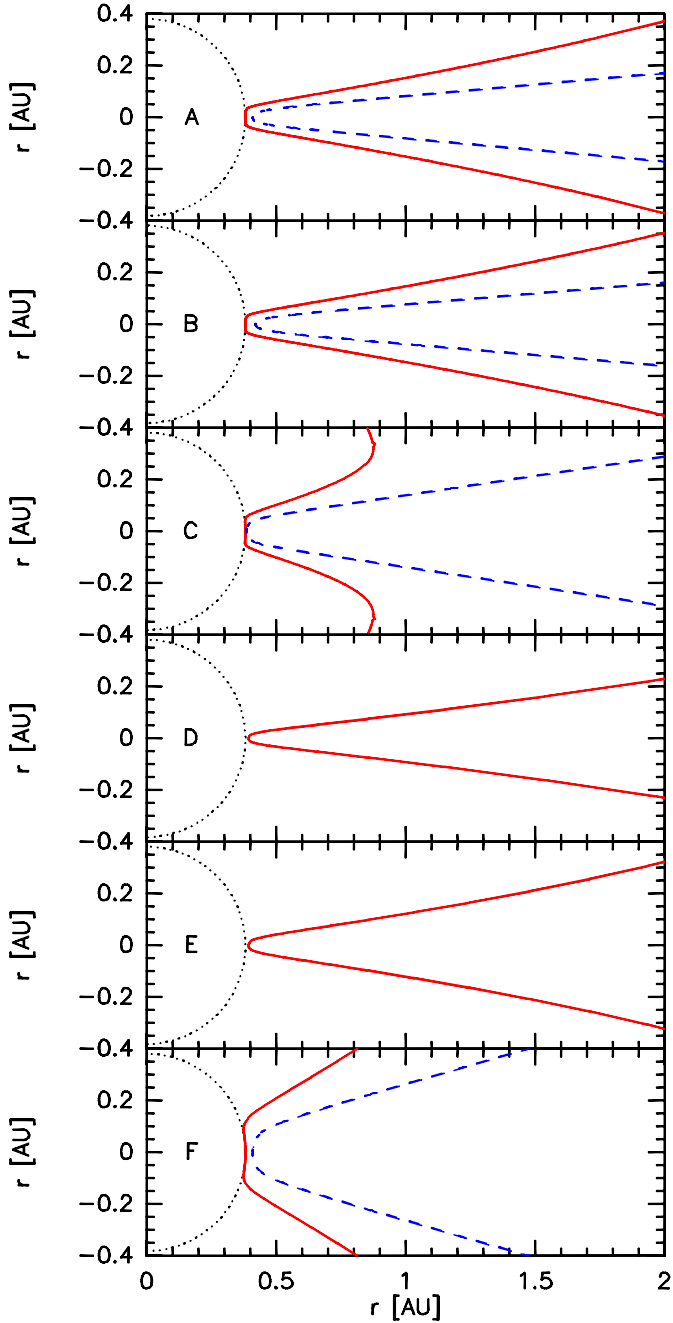}}
\caption{Ionization radii in equatorial regions of H (dashed lines) and He (solid
lines) for the different model calculations in the optically thin approximation. The stellar
surface is shown as the dotted line.}
\label{rs_duenn}
\end{figure}
 
\begin{figure}
\resizebox{\hsize}{!}{\includegraphics{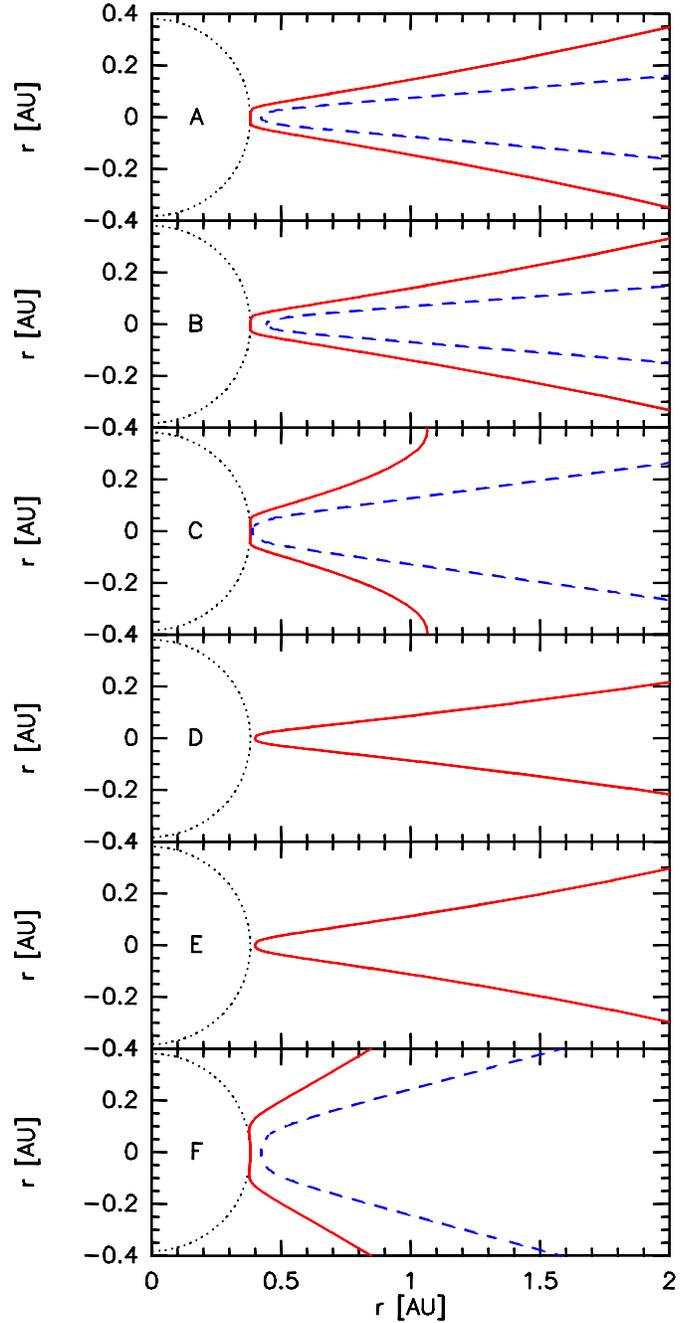}}
\caption{As in Fig.\,\protect{\ref{rs_duenn}} but for the on-the-spot approximation.}
\label{rs_ots}
\end{figure}    

\begin{figure}
\resizebox{\hsize}{!}{\includegraphics{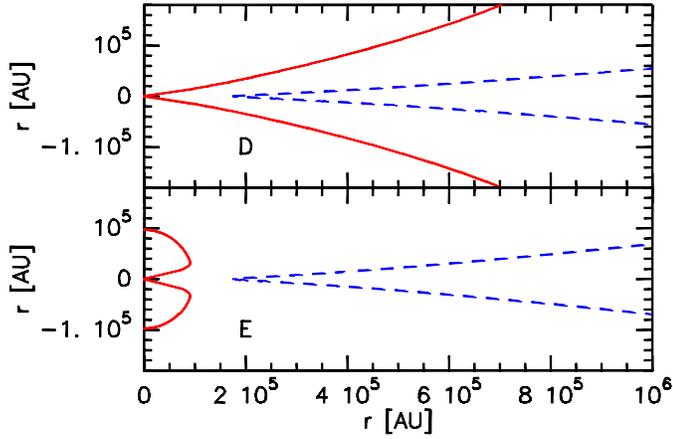}}
\caption{Optically thin approximation for the models D and E but for a much larger distance.
Shown are the ionization radii for H (dashed line) and He (solid line).}
\label{rs_duenn2}
\end{figure}

\begin{figure}
\resizebox{\hsize}{!}{\includegraphics{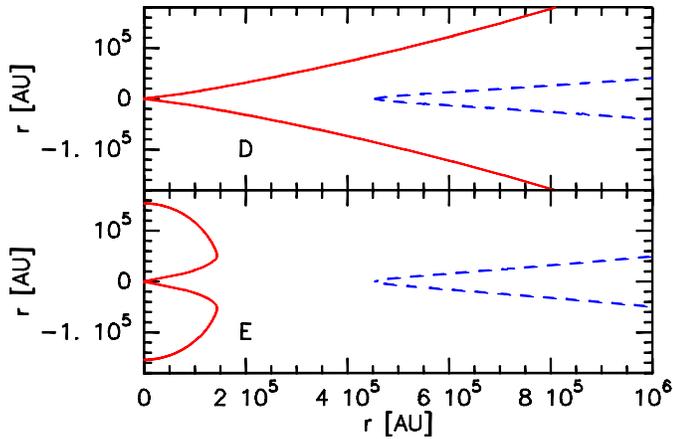}}
\caption{On-the-spot approximation for the models D and E but for a much larger distance.
Shown are the ionization radii for H (dashed line) and He (solid line).}
\label{rs_ots2}
\end{figure}

The ionization radii, $R_{\rm s}(\theta)$, calculated for the different model parameters (see 
Table\,\ref{windtab}), are presented in Fig.\,\ref{rs_duenn} for the optically thin approximation 
and in Fig.\,\ref{rs_ots} for the on-the-spot approximation. In addition, Figs.\,\ref{rs_duenn2} 
and \ref{rs_ots2} show the ionization radii of the models D and E but for a much larger distance. 

The most important result is that for all of the models there is a disk-shaped neutral region 
around the star, and for the high-density models (A, B, C, and F) these neutral disks 
can even exist at or just above the surface of the hot star.

Due to the higher number of ionizing photons available in the on-the-spot 
approximation we would for this method expect the ionization radii to be larger
compared with the optically thin results, because e.g. in the case of He the size of the 
ionization radius depends only on the recombination coefficient.
Comparing Figs.\,\ref{rs_duenn} and \ref{rs_ots} this effect is not obvious for the 
equatorial direction due to the choosen scaling. But  
the ionization radii calculated in the on-the-spot approximation are indeed larger
as can be seen by comparing the results for a more polar
direction or by comparing the Figs.\,\ref{rs_duenn2} and \ref{rs_ots2}. 
The real location of the ionization radius should lie in between the two boundaries 
defined by the optically thin and the optically thick approximation.

We take model A as the reference model and discuss now the effects of the change in the 
different free parameters on the size of the ionization radii. The higher electron 
temperature in model B leads to a slightly larger ionization radius because the 
recombination coefficient and therefore the number of recombinations decreases with 
increasing electron temperature (see Fig.\,\ref{rec}). The influence of the electron 
temperature is, however, not very significant.

The series of the models C, A, and D displays the behaviour of the ionization radii 
for a decreasing mass flux, i.e. a decreasing hydrogen density in equatorial direction. 
Consequently, the ionization radii of H and He increase drastically because (i) much less 
recombinations occur, and (ii) for a lower density, i.e. a lower optical depth in direction 
to the star, the ionizing photons can penetrate much deeper into the wind. While the density 
in model A is still high enough for effective recombination and for an effective absorption 
of the ionizing photons near the stellar surface, especially the H ionization radius in 
model D is shifted to a much larger distance (see Figs.\,\ref{rs_duenn2} and \ref{rs_ots2}).

Not only the change in mass flux leads to a change in density. Also a lower value of
$b$, the density ratio between equator and pole, results in a lower density and leads
to a shift of the ionization radius to a much larger distance (model E, see also
Figs.\,\ref{rs_duenn2} and \ref{rs_ots2}).
 
And finally, the smaller value of $s$ in model F, which means that the concentration of the
density to equatorial directions is less strong, leads to a much broader neutral disk-like
structure, as expected.

\section{Ionization structure of H and He}\label{ionstruc}

In the previous section we calculated the ionization structure using the approximation
that the wind is fully ionized at $r(\theta) < R_{\rm s}(\theta)$ and
neutral at $r(\theta) > R_{\rm s}(\theta)$. This is a severe simplification because the 
region where the wind is partly ionized may be substantial. Therefore, in this section we
calculate the ionization structure of the wind more exactly.
As concluded from the Figs.\,\ref{rs_duenn} and \ref{rs_ots}, the neutral region in 
equatorial direction can already be reached at or near the stellar surface, as long as the 
wind density is high enough.
For these high density models, the on-the-spot approximation is a very good tracer for the
H and He ionization radii and we use this approximation further on to calculate 
the ionization structure in these winds in equatorial directions.

We start with the ionization balance equations in the on-the-spot approximation 
derived above, i.e. Eq.\,(\ref{H_ots2}) for H and Eq.\,(\ref{He_ots2}) for He.
The optical depth is given by the Eqs.\,(\ref{tau1}) and (\ref{tau2}).
From Eqs.\,(\ref{H_ots2}) and (\ref{He_ots2}) the ionization fractions for H and He 
can be calculated radially outwards in the wind whereby the optical depth at each 
point in the wind between the stellar surface and the distance $r$ must be known.
The inclusion of the optical depth along the direction to the star into the 
calculations leads to results for the ionization radii which are more accurate than in
the approximate calculation of the previous section.    

\begin{figure}
\resizebox{\hsize}{!}{\includegraphics{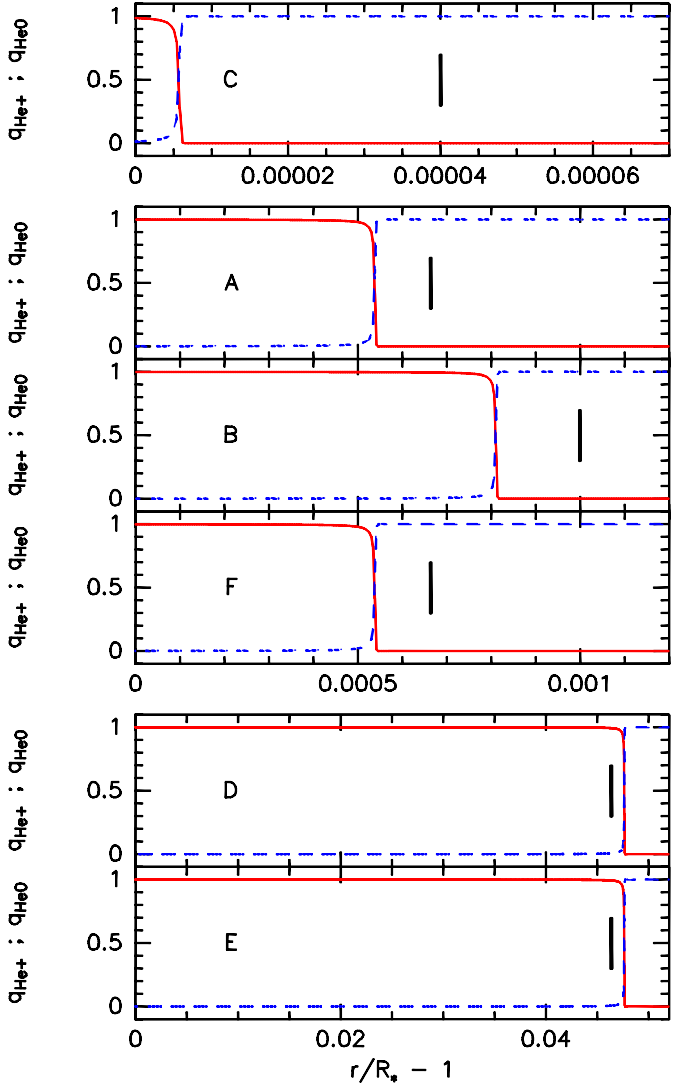}}
\caption{He ionization structure in equatorial direction ($\theta = 90\degr$) for the
different models. The models are grouped according to their density in equatorial direction:
the upper panel belongs to the highest density, the lower panel to the lowest density models.
The shift of the transition radius to higher distances for lower densities is clearly visible.
The solid lines represent $q_{\rm He^{+}}$ and the dashed lines
$q_{\rm He^{0}} = 1 - q_{\rm He^{+}}$. The thick vertical line is the location of
the ionization radius from Fig.\,\protect{\ref{rs_ots}}.}
\label{dens_He}
\end{figure}

In Fig.\,\ref{dens_He} we plot the fractions of ionized and neutral helium calculated
for the different models of Table\,\ref{windtab}. The plots in the three panels are thereby
grouped according to the values of the equatorial hydrogen density, i.e. model C
in the top panel has the highest density, while the models D and E in the bottom panel 
have the lowest density. Also shown, as a thick vertical line, are the locations of the 
approximate ionization radii extracted from Fig.\,\ref{rs_ots}. 

At first, in the top and middle panel the distance of the approximate 
ionization radius is much larger than for the more accurate ionization 
fraction calculations. The reason for this is the optical depth which is 
not really zero as assumed in the approximate  ionization radius calculations. 
Instead, there exists always a very small fraction of neutral atoms which 
results in a non-negligible attenuation of the stellar radiation field.  
For example, in the highest density model C where the He neutral region 
nearly coincides with the stellar radius, the approximate method results 
in an ionization radius that is about a factor of 8 larger. This strong 
descrepency can be understood, if we look at the fraction of neutral He 
which turns out to be a few times $10^{-3}$ at the stellar surface. Such 
a large neutral fraction results in a substantial optical depth already 
at the stellar surface.

\begin{figure}
\resizebox{\hsize}{!}{\includegraphics{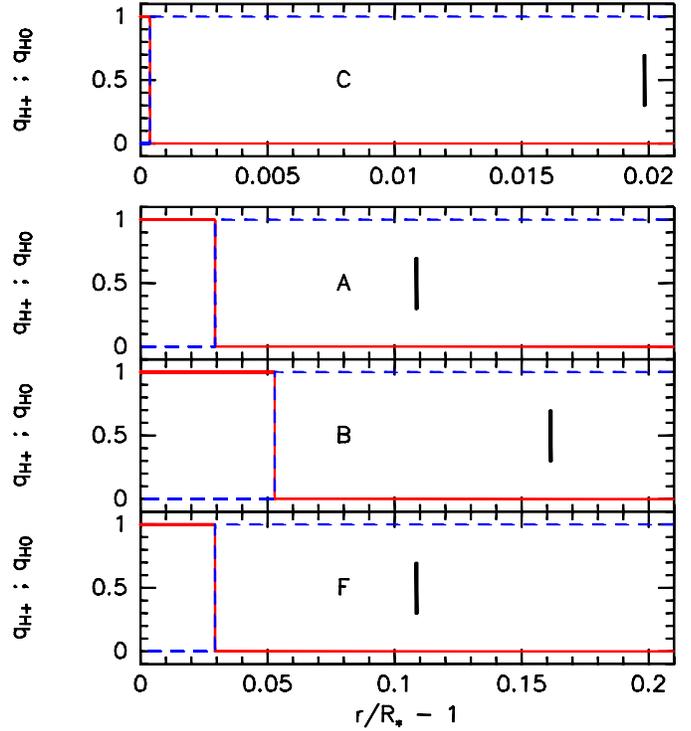}}
\caption{As Figure \protect{\ref{dens_He}} but for the H ionization structure. Models
D and E are not shown because due to their low densities no transition from ionized to
neutral occurs; the ionization fraction in these winds becomes constant.}
\label{dens_H}
\end{figure}

For the models A, B, and F, i.e. the models with the intermediate density, the He neutral 
region starts at larger radius and the neutral 
fraction inside this region is much smaller (only about $10^{-5}$--$10^{-6}$). The difference 
between the approximate and the real ionization radii is therefore much smaller,
about 25\%. 

In the lower panel of Fig.\,\ref{dens_He} the results for the models with the lowest density,
D and E, are shown. Here, the situation is different: the exact calculations lead to a larger 
distance of the neutral region than the ionization radius calculations. These seems not to 
be physically and shows us that in these two models the assumption of the validity of the 
on-the-spot approximation no longer holds because the density in the models D and E is 
already too low to guarantee that every photon produced by recombinations will be absorbed in 
the wind.

The main result of this comparison in Fig.\,\ref{dens_He} is that the approximate calculations, by neglecting the 
optical depth effects inside the ionized regions, overestimate the real locations of the 
ionization radii. The same behaviour can also be seen in Fig.\,\ref{dens_H} which displays 
the fractions of ionized and neutral hydrogen together with the locations of the approximate 
ionization radii.

Interestingly, when calculating the ionization fractions for the models D and E we could not
find the transition from ionized to neutral H, but instead the calculations resulted in
a constant ionization fraction contrary to what was found for the approximate calculations.
The same result was achieved for calculations in polar directions. This discrepancy can 
easily be understood by inspecting the balance equations (\ref{H_ots2}) and (\ref{He_ots2}): 
If we summarize on the right hand side the terms that contain the electron density, the
left hand side is proportional to $W(r)\cdot N_{\rm H}(r)$ while the right hand side
is proportional to $N_{\rm H}^{2}(r) \sim r^{-4}$. 
Since for large distances, the dilution factor becomes proportional to $r^{-2}$, the
left hand side also becomes proportional to $r^{-4}$ which then drops out of the equations.
The only $r$ dependent term that is still in is the optical depth. But for large
distances with only slowly varying $q$ values the optical depth becomes constant.
For large distances from the star, the ionization balance equations therefore become 
independent of $r$, and the ionization fractions become constant. As long as there is
no change in the density distribution, e.g. when the wind reaches the shockfront where the
wind material hits the interstellar medium, there will be no neutral region in the polar
directions of the wind but all material stays ionized with a constant ionization fraction.
In the approximate calculation of the ionization radii we make the assumption
that the neutral region starts at a distance smaller than infinity, but this assumption 
does not hold for low density winds with a $r^{-2}$ density 
distribution. Nevertheless, in winds with a constant 
density, as might be the case in planetary nebulae, the assumption is correct because
in a constant density wind the ionization balance equations never become independent of
$r$.

\section{Discussion}\label{discuss}

The motivation for this paper was to find the location of neutral material in the equatorial 
regions around hot and massive stars. In our model calculations we make some assumptions 
that seem not to be valid for the regions of our interest, i.e. very near the stellar
surface. In this section we therefore discuss these assumptions and show that they are indeed 
justifiable also near the stellar surface.

\begin{enumerate}

\item {\bf The constant electron temperature.} In a test calculation we computed with the help of 
the thermal equilibrium equations for a pure H plus He wind the electron temperature 
distribution in equatorial direction. We found that on this short distance between stellar 
surface and neutral region the electron temperature varies only very slightly, so that our 
assumption of a constant electron temperature seems to be valid, at least for the equatorial 
regions.  

\item {\bf The constant wind velocity.} The wind velocity of a hot star normally increases from a 
value $< 1$\,km/s in the stellar photosphere to the terminal velocity which is reached at
a distance of a few stellar radii. In polar direction, the B[e] supergiants show a typical
radiation-driven CAK-type wind with a terminal velocity of about 2000\,km/s. This value
was used in our calculations. Since the ionization radius in polar directions lies at
distances of $10^{5} - 10^{7}$\,AU, the assumption of a constant wind velocity, i.e. the 
terminal velocity in these directions, is valid. The only exception is model C which is the
wind of highest density in our sample. Here, the polar ionization radius is reached
already at a distance of 0.9\,AU. For this model a proper inclusion of the radial velocity 
distribution is necessary when the correct value of the polar ionization radius is needed.

In the equatorial direction, in which we are mainly interested, our choice of the terminal
velocity clearly overestimates the real velocity in these regions, since we found that
the ionization radii here are reached already very close to the stellar surface where the 
wind has not yet been accelerated effectively. However, if we would include the correct, 
very slow wind velocity, the hydrogen density in equatorial direction which is proportional
to $v^{-1}$ would be much larger than in our model calculations. The overestimation of the
wind velocity in our model therefore shifts the ionization radius to a larger distance.
 
\item {\bf The point source approximation} for the mean stellar radiation field.
For distances near the star we normally would
have to take into account effects like limb darkening and the size of the stellar disk.
Both effects lead to a decrease in the mean optical depth at distance $r$ from the star.
Ignoring these two effects, Eq.\,(\ref{stellar}) results in an overestimation of the stellar
radiation field and in a shift of the ionization radius in equatorial direction to a larger
distance.

\end{enumerate}

The assumption of a constant wind velocity that equals the terminal velocity
and the assumption of the point source approximation for the mean stellar
raditation field result both in a shift of the ionization radius to larger
distances from the star. Our calculations therefore reflect the ``worst case" and 
give an {\it upper limit} for the distance of the equatorial neutral region.

\section{Comparison with observations from other hot stars}

The B[e] supergiants are not the only class of stars that shows non-spherically symmetric 
winds and we mention here two galactic stars, the luminous blue variable (LBV) AG 
Carinae and the peculiar B[e] star MWC\,349\,A, that are also found to have emission from CO 
and dust coming from regions near the stellar surface.

AG Carinae is a LBV star, even one of the most luminous ones with $L = 1.7\cdot 
10^{6}\,L_{\sun}$ (Humphreys et al. \cite{humphreys}) and with a variable effective 
temperature between 9000\,K and 25\,000\,K (Lamers \cite{lamers86}; Voors et al. 
\cite{voors}). The mass of AG Carinae is of order 35\,$M_{\sun}$ (Vink \& de Koter 
\cite{vink}). It is surrounded by a very bright, extended nebula which is composed of ionized 
gas and dust. Most recently, the properties of the neutral material around this star was 
subject to a project of Nota et al. (\cite{nota}). The authors observed the lowest 
rotationally transitions of CO, and with their spatially resolved observations they found, 
most astonishingly, that the CO emission arises in regions near the star. The authors 
therefore suggested that the CO is located in an outflowing disk, just as it is the case for 
B[e] supergiants. Since LBV's have very high mass loss rates of order $10^{-5} - 
10^{-4}\,M_{\sun}$yr$^{-1}$ even at times in between individual outbursts (e.g. Leitherer et 
al. \cite{leitherer}), the wind scenario in AG Carinae might equal those of the B[e] 
supergiants, and a neutral disk extending nearly down to the stellar surface might surround 
AG Carinae. Our model calculations shown above speak in favour of such a scenario.

The other interesting star is the peculiar B[e] star MWC\,349\,A. Although known for nearly
100 years, its evolutionary state is still under debate because its spectra show 
characteristics of a pre-main sequence B[e]-type star as well as characteristics of a B[e] 
supergiant (Lamers et al. \cite{lamers98}). MWC\,349\,A has a luminosity of $3\cdot 
10^{4}\,L_{\sun}$, a mass loss rate of $1.2\cdot 10^{-5}\,M_{\sun}$yr$^{-1}$ (Cohen et al. 
\cite{cohen}), an effective temperatue of $35\,000$\,K, and a mass of 25 -- 30\,$M_{\sun}$ 
(Thum et al. \cite{thum92}; \cite{thum94}). This star is known to possess a Keplerian 
rotating disk. Strong CO first-overtone band emission has been observed which must be located 
near the star (Kraus et al. \cite{kraus2}), i.e. much nearer than about 3\,AU which is the 
dust evaporation radius (Kraus \cite{kraus1}).

These are only two examples that show how common the phenomenon of neutral gas nearby 
hot stars and located in a shielded, outflowing disk is. 

In our calculations presented in this paper we took an empirical density
distribution that accounts for the observed quantities like the density ratio 
between equator an poles, the mass loss rate and velocity in polar directions, 
and the density concentration towards the equator. The physical reason for such 
kind of density distribution is not yet fully understood. Since the B[e] 
supergiants are known to be fast rotators, a differential rotation of the star 
with a drastically  increase of rotation speed towards the equator
might lead to the observed angle dependent mass flux. There 
exist several theoretical models to describe the concentration or focussing of the 
material towards the equatorial region. The most promising are the rotationally 
induced bi-stability mechanism (see Lamers \& Pauldrach \cite{lamers91}; Pelupessy 
et al. \cite{pelupessy}) and wind compression leading either to a wind compressed 
disk (Bjorkman \& Cassinelli \cite{bjorkman}; Bjorkman \cite{bjorkman98}; Owocki et al. 
\cite{owocki94}, \cite{owocki96}, \cite{owocki98}) or to a wind compressed zone (e.g. 
Ignace et al. \cite{ignace}). Unfortunately, none of them is able to explain all the
observed properties of the polar and equatorial winds self-consistently.
A better understanding of the physics of these outflowing disks and of the ionization 
structure in non-spherically symmetric winds is therefore necessary.

\section{Conclusion}

Our main result is that the equatorial winds of B[e] supergiants, which are formed due 
to a high mass flux of low velocity, can become neutral very near the stellar surface. 
In these neutral disk-shaped regions, which are shielded from direct stellar
radiation, molecules and dust can form. The existence of molecules and dust in the 
disks around B[e] supergiants is already known from observations, but it is 
astonishing that our calculations result in neutral disks that extend down to the 
hot stellar surface. 

These results coincide very nicely with the most recent spatially resolved observations
of strong CO emission. This emission seems to originate in regions very near the hot 
LBV star AG Carinae. Its existence can only be understood if a neutral disk shields 
the material from stellar dissociation.
For this star, a scenario as predicted by our model calculations, seems   
to be a natural way to allow CO molecules to exist very close to the hot star.
The same holds also for the peculiar B[e] star MWC\,349. This star also shows
strong CO emission coming from the inner parts of a neutral Keplerian rotating disk. 

In addition we find that in polar directions the approximate ionization radius
calculation for winds with an $r^{-2}$ density distribution turns out to be wrong
because one uses the physically wrong assumption that there must be a turnover
from ionized to neutral at a distance smaller than infinity.
In fact, for large distances, the ionization balance
equations become independent of $r$ and the ionization fractions become constant
so that no neutral region can be found as long as the density in the wind does
not change, e.g. when reaching the shockfront where the wind hits the interstellar
medium. 

Although our model calculations are restricted to a simple hydrogen plus 
helium wind, they nevertheless mirror the general behaviour of the ionization
structure in non-spherically symmetric winds of hot stars with a high mass flux in 
equatorial direction. 
Moreover, since the stellar parameters in our calculations 
are for a very luminous B[e] supergiant, the results are valid for the complete
group of B[e] supergiants, because for less luminous stars the number of Lyman
continuum photons decreases so that the wind becomes neutral even closer to the star.

\begin{appendix}

\section{Derivation of the classical Str\"{o}mgren sphere}

To show that the classical Str\"{o}mgren sphere relation follows from our
more general formulae we start with Eq.\,(\ref{rs_H_thin}) for a pure hydrogen nebula
of constant density 
\begin{equation}
4\pi W(r)
(1-q(r))\int\limits_{\nu_{0}}^{\infty}\frac{B_{\nu}a_{\nu}}{h\nu}e^{-\tau_{\nu}}d\nu
= \alpha q(r) N_{\rm H} \label{A1}
\end{equation}
In addition we use the point source approximation, i.e. $W(r) =
1/4 (R_{*}/r)^{2}$, and multiply both sides of the
equation by $4\pi r^{2}$  
\begin{equation}
(1-q(r))\int\limits_{\nu_{0}}^{\infty}4\pi R_{*}^{2}\frac{\pi B_{\nu}a_{\nu}}{h\nu}
e^{-\tau_{\nu}}d\nu = \alpha q(r) N_{\rm H} 4\pi r^{2} \label{A2}
\end{equation}
Now, we integrate both sides over the radius and take into account the
condition for $q(r)$ which is defined by Eq.\,(\ref{q_define}). The optical depth
can then be calculated
\begin{equation}
\tau_{\nu}\!\!=\!\!\int\limits_{R_{*}}^{r}\!\!a_{\nu}N_{\rm H}(1\!-\!q(r))
dr\!=\!\int\limits_{R_{\rm S}}^{r}\!\!a_{\nu}N_{\rm H}dr\!=\!a_{\nu}N_{\rm 
H} (r\!-\!R_{\rm S})
\end{equation}
and Eq.\,(\ref{A2}) becomes
\begin{equation}
\int\limits_{\nu_{0}}^{\infty}\!\!4\pi R_{*}^{2}\frac{\pi B_{\nu}a_{\nu}}{h\nu}
\int\limits_{R_{\rm S}}^{\infty}\!\!e^{-a_{\nu}N_{\rm H} (r - R_{\rm S})} dr d\nu =
\alpha N_{\rm H} 4\pi \int\limits_{R_{*}}^{R_{\rm S}}\!\!dr 
\end{equation}
The integration over the radius on the left-hand side results simply in
\begin{equation}
\int\limits_{R_{\rm S}}^{\infty}e^{-a_{\nu}N_{\rm H} (r - R_{\rm S})} dr =
\frac{1}{a_{\nu}N_{\rm H}}
\end{equation}
so that we end up with
\begin{equation}
\int\limits_{\nu_{0}}^{\infty}4\pi R_{*}^{2}\frac{\pi B_{\nu}a_{\nu}}{h\nu}
\frac{1}{a_{\nu}N_{\rm H}}d\nu 
= \alpha N_{\rm H} \frac{4\pi}{3} (R_{\rm S}^{3} - R_{*}^{3})
\end{equation}
or, if we use $L_{\nu} = 4\pi R_{*}^{2} \pi B_{\nu}$ and the assumption
$R_{\rm S}\gg R_{*}$,
\begin{equation}
\int\limits_{\nu_{0}}^{\infty}\frac{L_{\nu}}{h\nu} d\nu = \alpha N_{\rm H}^{2}
\frac{4\pi}{3} R_{\rm S}^{3}
\end{equation}
which is the classical formula for the Str\"{o}mgren sphere in a constant
density nebula.

For a density distribution of the form $N_{\rm H}(r) = A\cdot r^{-2}$ the
Str\"{o}mgren sphere is  
\begin{equation}
\int\limits_{\nu_{0}}^{\infty}\frac{L_{\nu}}{h\nu}\left( 1-e^{-a_{\nu}A/R_{\rm S}}
\right) d\nu = \alpha A^{2} 4\pi \left(\frac{1}{R_{*}} - \frac{1}{R_{\rm S}}
\right)
\end{equation}
which can be derived from our general equations in a similar way.  

\end{appendix}

\begin{acknowledgements}

We thank the unknown referee for critical comments that have led to an improvement of 
this paper. This work was partially supported by the German 
\emph{Deut\-sche For\-schungs\-ge\-mein\-schaft, DFG\/} grant number Kr~2163/2--1.

\end{acknowledgements}

\end{document}